\documentclass{amos}

\usepackage{amsmath}
\usepackage{amssymb}
\usepackage{siunitx}
\usepackage{xcolor}
\usepackage[numbers]{natbib}
\usepackage[hidelinks]{hyperref}
\usepackage{mdframed}

\newcommand{\crumb}[1]{\begin{mdframed}[backgroundcolor=cyan!40]#1\end{mdframed
}}

\title{\TitleFont Light curve completion and forecasting using fast and
scalable Gaussian processes (MuyGPs)}

\author{
	Im\`ene R. Goumiri \\
	Physics Division,
	Lawrence Livermore National Laboratory 
\and\vspace{0.5em}
  Alec M. Dunton \\
  Center for Applied Scientific Computing,
  Lawrence Livermore National Laboratory 
\and
	Amanda L. Muyskens \\
	Engineering Division,
	Lawrence Livermore National Laboratory 
\and
	Benjamin W. Priest \\
	Center for Applied Scientific Computing,
	Lawrence Livermore National Laboratory 
\and
	Robert E. Armstrong \\
	Physics Division,
	Lawrence Livermore National Laboratory 
}

\date{}

\begin{document}

\maketitle

\vspace{-0.8in}
\begin{center}
  LLNL-PROC-839253
\end{center}
\vspace{0.5in}

\begin{abstract}\normalsize

  Temporal variations of apparent magnitude, called light curves, are
  observational statistics of interest captured by telescopes over long periods
  of time.
  Light curves afford the exploration of Space Domain Awareness (SDA)
  objectives such as object identification or pose estimation as latent
  variable
  inference problems.
  Ground-based observations from commercial off the shelf (COTS) cameras remain
  inexpensive compared to higher precision instruments, however, limited sensor
  availability combined with noisier observations can produce gappy time-series
  data that can be difficult to model.
  These external factors confound the automated exploitation of light curves,
  which makes light curve prediction and extrapolation a crucial problem for
  applications.
  Traditionally, image or time-series completion problems have been approached
  with diffusion-based or exemplar-based methods.
  More recently, Deep Neural Networks (DNNs) have become the tool of choice due
  to their empirical success at learning complex nonlinear embeddings.
  However, DNNs often require large training data that are not necessarily
  available when looking at unique features of a light curve of a single
  satellite.

  In this paper, we present a novel approach to predicting missing and future
  data points of light curves using Gaussian Processes (GPs).
  GPs are non-linear probabilistic models that infer posterior distributions
  over functions and naturally quantify uncertainty.
  However, the cubic scaling of GP inference and training is a major barrier to
  their adoption in applications.
  In particular, a single light curve can feature hundreds of thousands of
  observations, which is well beyond the practical realization limits of a
  conventional GP on a single machine.
  Consequently, we employ MuyGPs, a scalable framework for hyperparameter
  estimation of GP models that uses nearest neighbors sparsification and local
  cross-validation.
  MuyGPs allows us to quickly train models on light curve data in seconds on a
  single workstation.
  In this manuscript we explore light curve completion and forecasting, and
  compare embeddings of the light curve data into multi-dimensional spaces to
  take advantage of daily and yearly periodicity.
  We show that our method outperforms feedforward DNNs both in terms of
  accuracy and quantity of training data required.

\end{abstract}

\section{Introduction}

Photometric light curves are a series of observations that track the brightness
of an object over a period of time.
They can be used to characterize the
dynamic properties of a system.
Astronomers frequently analyze light curves to understand a
whole range of astrophysical topics including: detailed physics within a
star~\cite{gaia2019}, the discovery of exoplanets~\cite{kepler2013},
classifying distant supernovae~\cite{des2022}, and characterizing the
population of near earth asteroids~\cite{atlas2018}.

Space Domain Awareness (SDA) involves monitoring, detecting and understanding
the population of earth-orbiting bodies or resident space objects (RSOs).
SDA is of increasing importance due to the incipient growth in the number of
space objects and debris orbiting the earth, due in large part to the recent
commercial development of satellites.
The number so such RSOs is expected to grow by orders of magnitude in the next
decade.
While SDA is concerned with maintaining custody of orbiting objects, it also
prioritizes the rapid identification of changes to and anomalous behavior of
the many varying orbital systems.

Fortunately, light curves are valuable for inspecting RSOs in much the same way
as astrophysical phenomena, and can enable critical information for SDA.
The brightness of an RSO depends on structural features like size, shape, and
material composition as well the geometry between the object, the sun and
observer.
The proliferation of low cost commercial-off-the-shelf (COTS) ground-based
telescopes has made the generation of light curves from RSOs easier to produce.
Furthermore, many constellations of ground-based telescopes have been tasked
with tracking RSOs for SDA.
These factors have enabled the relatively cheap production of large volumes of
RSO
light curves, which has prompted practitioners to apply automation to analyze
them at scale.

Practitioners have recently applied machine learning to light curves of RSOs to
solve various SDA tasks.
For example, comparing the light curve of an unknown RSO to a catalog of known
RSOs is useful for predicting features like
shape~\cite{linares2014space, furfaro2019shape}, material
composition~\cite{dianetti2019space}, and general categories or
classes~\cite{linares2016space, jia2018space, furfaro2018space}.
Furthermore, forecasting RSO light curves into the future is useful for
detecting deviations from the expected patterns-of-life in near-real-time,
affording the detection of anomalous events such as
maneuvers~\cite{shabarekh2016novel, dupree2021time} or configuration changes.

The goal of machine learning in this context is to learn a function mapping an
input space (typically the time domain) to a an observation space, e.g.
magnitude, based upon independent and identically distributed (i.i.d.)
samples from a distribution of input-observation pairs.
We will refer to this data distribution as the ``target distribution''.
A machine learning model is successful when it is able to accurately predict
the observation of an unseen input drawn from the target distribution.

Both diffusion-based or exemplar-based methods have been traditionally
used in image or time-series completion problems.
Diffusion-based methods~\cite{sohl-dickstein2015deep} use thermal diffusion
equations to propagate information from surrounding regions into missing
regions, they are mostly effective when the gaps are small and tend to smooth
details out for larger problems.
Exemplar-based methods~\cite{criminisi2004region} use greedy algorithms
to apply patches of training data to missing regions which can produce
implausible patterns.

Deep Neural Networks (DNNs) are an especially popular tool to model light
curves in the literature due to their expressiveness and generalization
capabilities.
A DNN is a type of representation learning model --- a machine learning model
that learns an appropriate feature representation of the data in addition to
producing predictions.
DNNs iteratively transform the input space into latent feature spaces using
linear functions that are ``activated'' by element-wise nonlinear functions.
DNNs also happen to be universal function approximators --- it is possible to
approximate any continuous function to an arbitrary level of precision using a
sufficiently large DNN.
The persistent popularity of DNNs derives from several sources, such as the
widespread availability of hardware accelerators like graphical processing
units (GPUs),  advanced stochastic optimization techniques that aid in their
training, and the development of user-friendly software libraries such as
Tensorflow~\cite{Abadi_TensorFlow_Large-scale_machine_2015} and
PyTorch~\cite{NEURIPS2019_9015}.

Although DNNs have many positive features, they also have some drawbacks that
are particularly notable in the light curve problem.
First, modern DNN architectures typically consist of a very large number of
parameters that must be trained.
Training such models generally consumes a large amount of computing resources,
as the model and its gradient are evaluated over many iterations in order to
refine parameter values.
In addition to computational expense, training a large model typically requires
a large amount of data.
The literature has observed a roughly linear relationship between model size
and the amount of labelled data required to train it~\cite{tan2018survey}.
This means that learning an accurate light curve model can require a large
number of observations in general.

Second, while DNN models lend themselves to high dimensional feature spaces
they tend to struggle with very small numbers of dimensions.
Feature engineering can often solve this problem, especially in time series
scenarios where some notion of a moving window usually serves as a feature
space.
However, this strategy is most successful when the observation cadence of the
light curve is high.
Upcoming ground-based surveys of deep space, such as the Legacy Survey of Space
and Time (LSST) are expected to incidentally capture many RSO images from which
light curves can be extracted.
However, these light curves will be irregular, sparse, and have low-dimensional
feature spaces, increasing the challenge of applying existing techniques.

Third, effective training typically requires a large volume of training data
that is complete and representative of the target distribution
i.e. a large model requires a large amount of training data that is i.i.d.
according to the target distribution.
In addition to informing prediction accuracy, data independence and size helps
complex DNN models avoid overfitting and simply memorizing the training data.
However, collections of physical measurements are often limited by the
realities of sensor availability and physical obstruction.
For example, weather affects optical astronomical measurements by introducing
correlated noise or blocking the desired object from view entirely.
Furthermore, it is in general desirable to design a model that can be
alternatively applied to several different RSO inference problems, i.e. several
distinct target distributions.
A transfer learning approach-one where the model is at least partially trained
on data from a possibly different distribution than the target
distribution~\cite{tan2018survey}- could address this need for a large volume
of training data.
Indeed, transfer learning has been applied to satellite aperture radar
images~\cite{rostami2019deep}, radiofrequency interference
characterization~\cite{lefcourt2022space}, and classifying RSOs using light
curves~\cite{furfaro2018space}.
However, transfer learning ultimately relies on features learned from a
potentially unrelated dataset, which may lead to inefficient or inaccurate
conclusions.

Generative models such as Variational Autoencoders (VAEs) and Generative
Adversarial Networks (GANs) are additional alternatives from the machine
learning literature that address data efficiency.
VAEs are autoencoders that impose structure upon their learned encoder and
decoder functions to ensure that the latent representation of the data encodes
the target distribution.
Researchers have successfully applied VAEs to learn shapes from the light
curves of RSOs~\cite{furfaro2019shape}.
GANs conversely simultaneously train models that generate samples meant to
mimic the target distribution while distinguishing between real and synthetic
samples.
Practitioners have recently used GANs to classify stars based upon their light
curves~\cite{garcia2022improving}.
However, GANs can be challenging and expensive to train.
GANs infer a distribution from a very small training dataset
and can suffer from mode collapse, non-convergence or general instability.

This brings us to the fourth point --- DNNs have difficulty quantifying the
uncertainty in their predictions.
It is generally difficult to determine when a DNN is extrapolating ---
predicting the response on data from a different distribution or in a different
region of the training data.
This is problematic in scientific applications because overconfident
predictions can lead to incorrect inference and unnatural results that are
difficult to interpret.
Furthermore, it is important for decision makers to distinguish between low and
high-confidence predictions when drawing conclusions for SDA.

Although uncertainty quantification methods are in research, most practical
models employed by practitioners provide only point predictions with no obvious
mechanism to measure model confidence.
Recent attempts to provide uncertainty quantification to DNNs attempt to learn
prediction intervals (i.e. error bars) in addition to point predictions, but
this literature is still developing and there is as yet no consensus
solution~\cite{kabir2018neural}.
Others attempt a hybrid approach where a Gaussian process is appended to the
last layer of a DNN~\cite{bradshaw2017adversarial}.

In this manuscript we propose Gaussian Process (GP) models as alternatives for
the light curve modeling problem.
Like DNNs, GPs are representation learning methods that are data-driven
universal function approximators.
A GP is a type of kernel method that implicitly and non-linearly maps input
features into a possibly infinitely dimensional inner product space.
Kernel methods use a parameterized kernel function to cheaply compute inner
products in this implicit space to predict unknown responses.
GPs can also be thought of as the generalization of a multivariate normal
distribution, where all data within a defined domain is assumed to be jointly
normally distributed with the covariance defined by the kernel function.
Therefore, predictions from GP models are probabilistically defined through
conditional probability.
GPs are attractive for scientific applications due to this inherently Bayesian
inference model, which produces explicit uncertainty quantification (Gaussian
distributions) of its predictions.
Furthermore, GPs have been shown to outperform DNNs in data-starved
regimes~\cite{muyskens2022star} and are ideal models for low-dimensional
feature spaces~\cite{muyskens2021muygps}.

However, conventional GPs have very poor scaling in the number of training
observations, which has previously limited their application to the light curve
modeling problem.
Both realizing the predictions from the GP model and evaluating the likelihood
function in training require cubic computation in the number of data points,
and require quadratic memory to store the kernel matrix.
While this practical drawback has tended to limit the application of GPs to
small data problems, scalable approximate GP methods have proliferated in the
literature~\cite{heaton2019case, liu2020gaussian}.
These methods typical tradeoff accuracy for speed.
However, the approximate GP method MuyGPs offers a selection of settings that
demonstrates superior accuracy or computational scaling in several datasets
~\cite{muyskens2021muygps, muyskens2022star, buchanan2022gaussian}.

Therefore, in this paper we use the MuyGPs approximate GP estimation
method~\cite{muyskens2021muygps}.
MuyGPs uses nearest neighbors sparsification and a batched leave-one-out
cross-validation objective function to train a GP-like model in nearly linear
time in the number of data observations.
Investigators have successfully applied MuyGPs to cosmology image processing
problems~\cite{goumiri2020star, muyskens2022star, buchanan2022gaussian}.

In this paper, we propose a method of light curve interpolation and prediction
using MuyGPs that gives both predictions and uncertainty quantification of
those predictions.
We use this method in two modes:
\begin{enumerate}
  \item Interpolation, the prediction of unobserved magnitudes in the past, and
  \item Forecasting, the prediction of unobserved future magnitudes.
\end{enumerate}
Interpolation is useful as a data preprocessing utility for catalogs.
Interpolation allows us to fill in magnitude observations for sparse, irregular
light curves to make them suitable for downstream comparison tasks such as
shape, pose, size, type, or material estimation.
Forecasting is useful to detect deviations from expected future
patterns-of-life that correspond to anomalies.
The posterior variance provided by GPs is especially useful to determine what
constitutes a significant deviation from expected behavior.

In \autoref{sec:method}, we provide background on the light curve data itself,
data processing, as well as the machine learning methods we utilize in our
comparative study.
Then in \autoref{sec:res} we describe several numerical studies that
demonstrate the comparative performance of several choices of our GP method to
that of a DNN and how the uncertainty quantification provided by our method can
be used to detect anomalies.
Finally, in \autoref{sec:conclusion}, we discuss our conclusions, the
importance of our findings, limitations, and future work associated with our
methodology.

\section{Methodology}
\label{sec:method}
\subsection{Light curves definition}

\begin{figure}[htbp]
  \centering
  \includegraphics[height=\linewidth,
    angle=-90]{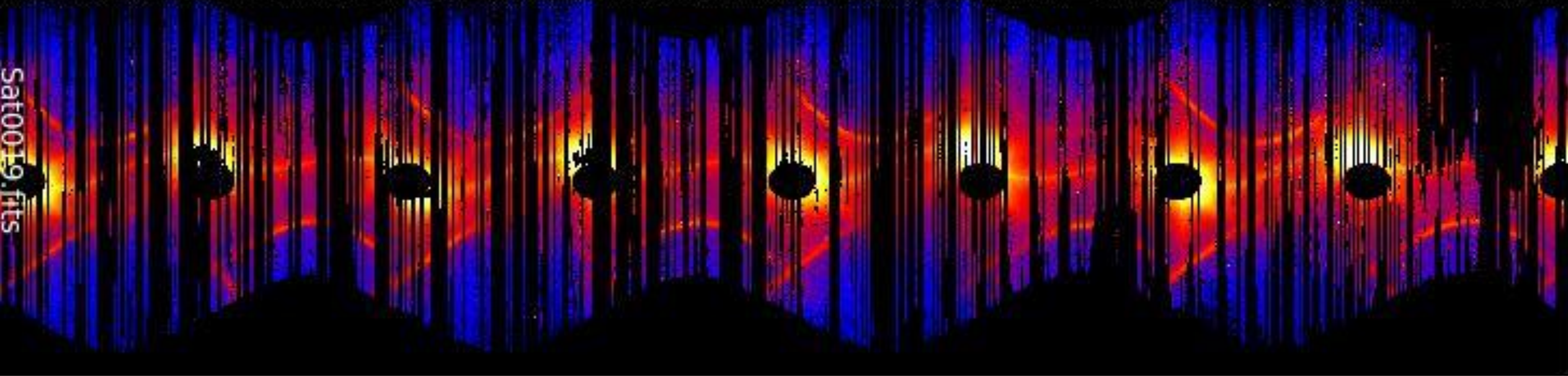}
  \caption{
    The light curve of a single object over 4 years.
    Each column represents a single day.
    The color scale represents the brightness.
    The black stripes and \emph{holes} highlight the sparsity of observations;
    the former are due to weather, the latter to eclipse.
    The seasonality and yearly periodicity induces the outer shape.
    Courtesy of~\cite{monetgrams}.
  }\label{fig:illustration:light_curve}
\end{figure}

Light curves are time series of the brightness of resident space objects over
long periods.
They are obtained by observing the same object, night after night, for several
days or even years.
One way to visualize them is in two dimensions with one axis representing the
time of day (or solar phase angle) and the other axis representing the days
while the pseudo-color indicates the brightness, as in
\autoref{fig:illustration:light_curve}.

Light curves of RSOs contain potentially detailed information.
For example, one could detect dust accumulating on highly reflective solar
panels and deduce the age and life expectancy of a satellite.
Higher intensity reflections also have the potential to inform about the shape
of the observed object since different facets of a multi-faceted object would
reflect differently.
In addition, satellites can maneuver or rotate, in which case, the reflected
light will deviate from the expected pattern.
Those deviations can be detected and increase SDA knowledge.

\begin{figure}[htbp]
  \centering
  a) \raisebox{-0.85\height}{\includegraphics[width=0.1\linewidth,
      height=0.1\linewidth]{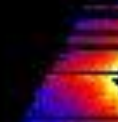}}
  b) \raisebox{-0.85\height}{\includegraphics[width=0.1\linewidth,
      height=0.1\linewidth]{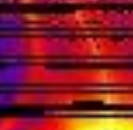}}
  c) \raisebox{-0.85\height}{\includegraphics[width=0.1\linewidth,
      height=0.1\linewidth]{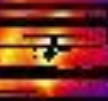}}
  d) \raisebox{-0.85\height}{\includegraphics[width=0.1\linewidth,
      height=0.1\linewidth]{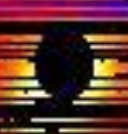}}
  \caption{
    Missing data due to a) time of day, b) weather, c) sensor saturation, and
    d) eclipse.
  }\label{fig:illustration:missing-data}
\end{figure}

Observations can be limited by the time of day, weather, sensor saturation, and
eclipses (see \autoref{fig:illustration:missing-data}).
Data is only available at night, and the measurements are less accurate near
dusk and dawn.
The weather can preclude observations for hours and sometimes days at a time.
When the reflected light from the sun is particularly bright, sensors can
saturate leading to missing data, though this particular case is usually fairly
obvious when looking at the brightness of surrounding available data.
Lastly, periodically, the light from a satellite will be eclipsed by the earth.

To test our codes we use a light curve dataset provided by Dave Monet
\cite{monetgrams} of 43 satellites.
All satellites in the dataset are from the public catalog
(www.space-track.org).
We selected 13 distinct RSOs, those manually flagged by Dave Monet as
\emph{nominal}, for our analysis.
There are approximately 500,000 data points per object.
All observations were taken from a single camera in Flagstaff AZ between
September 2014 and September 2018.
The dataset contains the brightness (magnitude; band is approximately Johnson
V) as well as the measurement error in brightness (uncertainty).

\subsection{Processing}\label{sec:processing}


To test our prediction capabilities, we select a portion of each light curve as
test data.
We use time periods ranging from a few hours (3 hours) to multiple weeks (2
weeks).
And we either select the test data at a random time within the time series
(interpolation) or at the end (extrapolation).
To guarantee that the selected period does not fall at a time lacking
observations (e.g. during the day or during a cloudy night) we reject those
that have fewer than 90\% of the data we'd expect to find in that time interval
if the data was uniformly distributed.
In addition, we apply that same condition to the preceding interval to ensure
that we're not accidentally interpolating for longer that we would expect.


Since the light curves have some daily and yearly periodicity, we compare
several embeddings of the light curve data into multi-dimensional spaces.
The reference is the 1D model which is just the original time series.
The 2D model has one dimension representing times of day as real numbers in the
$[0, 1)$ interval, and another dimension representing days as integers,
starting at zero on the first day of observation.
The 3D model is like the 2D model but with an additional dimension for years,
as integers between 0 and 4, and with a modified days dimension modulo 365.
Note that all the input dimensions are eventually normalized to the interval
$[0, 1]$ as is customary.

An alternate 2D model with year and time of year as opposed to day and time of
day was considered but dismissed for simplicity and for not being as intuitive
and interesting.
We observed that day-to-day correlation is stronger than year-to-year in our
dataset.

Note that it is more traditional to model this type of data periodicity using a
periodic kernel, but the MuyGPs estimation framework depends on a kernel
sparsity that is ultimately not validated with that kernel type.
Therefore, these embeddings are a novel way to replace such a periodic kernel
while maintaining the computationally efficient framework.

\subsection{Gaussian processes}

We will consider throughout a light curve to be a univariate response
$Y : \mathcal{X} \rightarrow \mathbb{R}$, where
$\mathcal{X} \subseteq \mathbb{R}^d$
is the observation space along $d$ time dimensions.
In preprocessing we de-trend the data, and so $Y$ therefore has zero mean.
In modeling $Y$ with a GP, we assume that it is drawn from a Gaussian process
distribution.
This means that $Y$ follows a multivariate Gaussian distribution at any finite
set of $n$ points $X = (\mathbf{x}_1, \dots, \mathbf{x}_n) \in \mathcal{X}^n$.
However, in reality measurement noise perturbs our observed values of $Y$ at
locations $X$.
We assume that each measurement is perturbed by homoscedastic noise that are
i.i.d.
$\mathcal{N}(0, \epsilon)$.
That is,
\begin{align} \label{eq:gp_prior}
  Y(X)
   & = (Y(\mathbf{x}_1), \dots, Y(\mathbf{x}_n))^\top \sim \mathcal{N}
  \left ( \widetilde{0}, Q_\theta(X, X) \right ),                      \\
  Q_\theta(X, X)
   & = \sigma^2 \left ( K_\theta(X, X) + \epsilon I_n \right ),
\end{align}
where $\mathcal{N}$ is the multivariate Gaussian distribution, $\widetilde{0}$
is the $n$-dimensional zero vector, $\sigma^2$ is a variance scaling term,
$I_n$ is the $n \times n$ identity matrix, and $K_\theta(X, X)$ is an
$n \times n$ positive definite, symmetric covariance matrix between the
elements of $X$ that is controlled non-linearly through kernel function
$K_\theta(\cdot, \cdot)$ with hyperparameters $\theta$.
$Q_\theta(X, X)$ is $K_\theta(X, X)$ that is perturbed by the measurement noise
prior $\epsilon$ and scaling parameter $\sigma^2$.

Similarly, any additional (possibly unobserved) datum
$\mathbf{x}^* \in \mathcal{X}$ is also jointly normal with observed data $X$ by
the GP assumption.
Thus, the conditional distribution for the response of $Y$ at $\mathbf{x}^*$
given responses observed at $X$ and $\theta$ is also multivariate normal with
mean and variance
\begin{align}
  \label{predmean}
  Y_\theta(\mathbf{x}^* \mid X)
   & =
  K_\theta(\mathbf{x}^*, X) Q_\theta(X, X)^{-1} Y(X), \text{ and}
  \\
  \label{predvar}
  \text{Var}(Y_\theta(\mathbf{x}^* \mid X))
   & =
  K_\theta(\mathbf{x}^*, \mathbf{x}^*) - K_\theta(\mathbf{x}^*, X)
  Q_\theta(X, X)^{-1} K_\theta(X, \mathbf{x}^*),
\end{align}
where $K_\theta(\mathbf{x}^*, X) = K_\theta(X, \mathbf{x}^*)^\top$ is the
cross-covariance matrix between $\mathbf{x}^*$ and the elements of $X$.

GPs are typically trained by maximizing the log-likelihood of the observations
$Y(X)$ with respect to $\theta$.
This log-likelihood function possesses the following form:
\begin{equation}
  \label{ll}
  log(L(\theta, Y(X))) = - \frac{p}{2}log(2 \pi)	- \frac{1}{2}
  log(|Q_\theta(X, X)|)  - \frac{1}{2} Y(X)^T Q_\theta(X, X)^{-1} Y(X).
\end{equation}
However, evaluating \autoref{ll} is $O(n^3)$ computation and $O(n^2)$ in
memory, which is intractable for all but relatively small data.
A MuyGPs model rewrites equations~\ref{predmean}~and~\ref{predvar} as
\begin{align}
  \label{muygpspred}
  \widehat{Y}_\theta(\mathbf{x}^* \mid X_{N^*})
   & =
  K_\theta(\mathbf{x}^*, X_{N^*}) Q_\theta(X_{N^*}, X_{N^*})^{-1} Y(X_{N^*}),
  \textrm{ and}
  \\
  \label{muygpsvar}
  \text{Var}(\widehat{Y}_\theta(\mathbf{x}^* \mid X_{N^*}))
   & =
  K_\theta(\mathbf{x}^*, \mathbf{x}^*) - K_\theta(\mathbf{x}^*, X_{N^*})
  Q_\theta(X_{N^*}, X_{N^*})^{-1} K_\theta(X_{N^*}, \mathbf{x}^*),
\end{align}
where $X_{N^*}$ are the nearest neighbors of $\mathbf{x}^*$ in
$X \setminus{\{\mathbf{x}^*\}}$.
MuyGPs trains $\theta$ in terms of some loss function $\ell(\cdot, \cdot)$ over
a sampled batch $B = (\mathbf{x}_1^*, \dots, \mathbf{x}_b^*) \subseteq X$ of
size $b$ by minimizing an objective function $Q(\theta)$.
Minimizing $Q(\theta)$ allows us to train $\theta$ without evaluating the
expensive determinant in the GP likelihood.
When we set $Q(\theta)$ to leave-one-out cross-validation and $\ell_\theta$ to
mean squared error, training $\theta$ reduces to minimizing the function
\begin{equation} \label{eq:batch_loss}
  Q(\theta)
  =
  \frac{1}{b} \sum_{i \in B} \left (
  Y(\mathbf{x}_i^*) - \widehat{Y}_\theta(\mathbf{x}_i^* \vert X_{N_i^*})
  \right )^2.
\end{equation}
Note other loss functions can be employed in this framework, but this mean
squared error function has been demonstrated as
performant~\cite{muyskens2021muygps}.
We use Bayesian optimization to optimize \autoref{eq:batch_loss} to train
$\theta$ in our experiments.

\subsection{A standard neural network model for benchmarking}

Neural networks have achieved state-of-the-art results on common benchmark
problems and are now ubiquitous in scientific applications.
Although the focal point of this work is the application of Gaussian processes
to light curve modeling, we benchmark the predictive performance of MuyGPs
against a standard deep neural network model to ensure its predictive accuracy.
We select a fully-connected architecture, as the structure of the feature space
is not suited to convolutional models.
Moreover, recurrent neural networks, including LSTMs, are another popular class
of models that are useful in forecasting settings, but not interpolation
problems.

Fully-connected neural networks can be represented as the composition of a
series of affine transformations and nonlinear activation functions.
The composition of an individual affine transformation and nonlinear activation
function constitutes a layer of the fully-connected network.
Each affine transformation in the network is parameterized by weights
$\mathbf{W}_i \in \mathbb{R}^{n_i \times n_{i+1}}$, where $n_i$ is the
dimension of the input to the $i^{th}$ layer and $n_{i+1}$ is the output
dimension, and a bias vector $\mathbf{b}_i \in \mathbb{R}^{n_{i+1}}$.
We label the activation function in the $i^{th}$ layer $\sigma_i$.
Let $\mathcal{F}_{NN}(\mathbf{x}) : \mathbb{R}^{n_{in}} \rightarrow
  \mathbb{R}^{n_{out}}$ be a neural network.
Then,
\begin{equation} \label{eq:FCNN}
  \mathcal{F}_{NN}(\mathbf{x})
  =
  \sigma_n \left(
  \mathbf{W}_n \sigma_{n-1} \left(
    \mathbf{W}_{n-1} \sigma_{n-2} \left(
      \mathbf{W}_{n-2} (\cdots) +\mathbf{b}_{n-2}
      \right) + \mathbf{b}_{n-1}
    \right) + \mathbf{b}_n
  \right).
\end{equation}

We train a fully-connected neural network using the day of the year and time of
day as the independent variables (features), with the target set to the
magnitude (normalized to take on values between 0 and 1 by dividing by the
largest magnitude observed).
The model architecture features ReLU activation functions and 5 layers of sizes
$2 \times 200$, $200 \times 200$, $200 \times 100$, $100 \times 20$, and $20
  \times 1$.
There are $62,420$ total parameters in the model, roughly $1/8-1/5$ the number
of training samples depending on the test case.
We train the neural network in batches of size 128 using the Adam optimizer for
100 epochs.
We use the mean-squared error loss function with 20 percent of the training
data withheld for validation, the training rate set to $10^{-3}$, and a
learning rate exponential decay factor of 0.5 applied every 30 training epochs.
We set the ratio of the number of parameters in the model to the number of data
points on the order of $1/10$, a commonly accepted ratio in the deep learning
community.
We selected these training hyperparameters based on common values used in deep
learning to achieve the best combination of training loss decay and validation
error minimization.

\section{Results}
\label{sec:res}

First we compare the predicting power of Gaussian processes for the different
embeddings of the input data described in \autoref{sec:processing}.
For each of the 13 light curves in our dataset, we randomly select a total of
20 time intervals, 5 for each time duration in \{3 hours, 1 day, 1 week, 2
weeks\}, and use those intervals as test data while using the rest as training
data.
The accuracy of the prediction is measured as the Root Mean Square Error (RMSE)
divided by the extent (max - min) of the magnitude in the test data.

We use the common radial basis function (or RBF) to define our kernel in all of
our experiments.
This means that we use the kernel function
\begin{equation} \label{eq:rbf}
  K_\ell(x, x^\prime)
  = \exp \left (
  -\frac{\|x - x^\prime\|_2^2}{2\ell^2}
  \right ).
\end{equation}
The RBF kernel has length scale parameter $\ell$, and our model additionally
has the measurement noise variance prior $\epsilon$ and variance scaling
parameter $\sigma^2$.
We fix $\epsilon = 10^{-5}$ in our experiments.
We train $\ell$ by optimizing equation \autoref{eq:batch_loss} via Bayesian
optimization.
However, no prediction-based objective function is sensitive to $\sigma^2$,
and so we optimize it according to the analytic procedure described in Section
2 of \citep{muyskens2021muygps}.

\begin{figure}[htbp]
  \centering
  \includegraphics[width=0.8\linewidth]{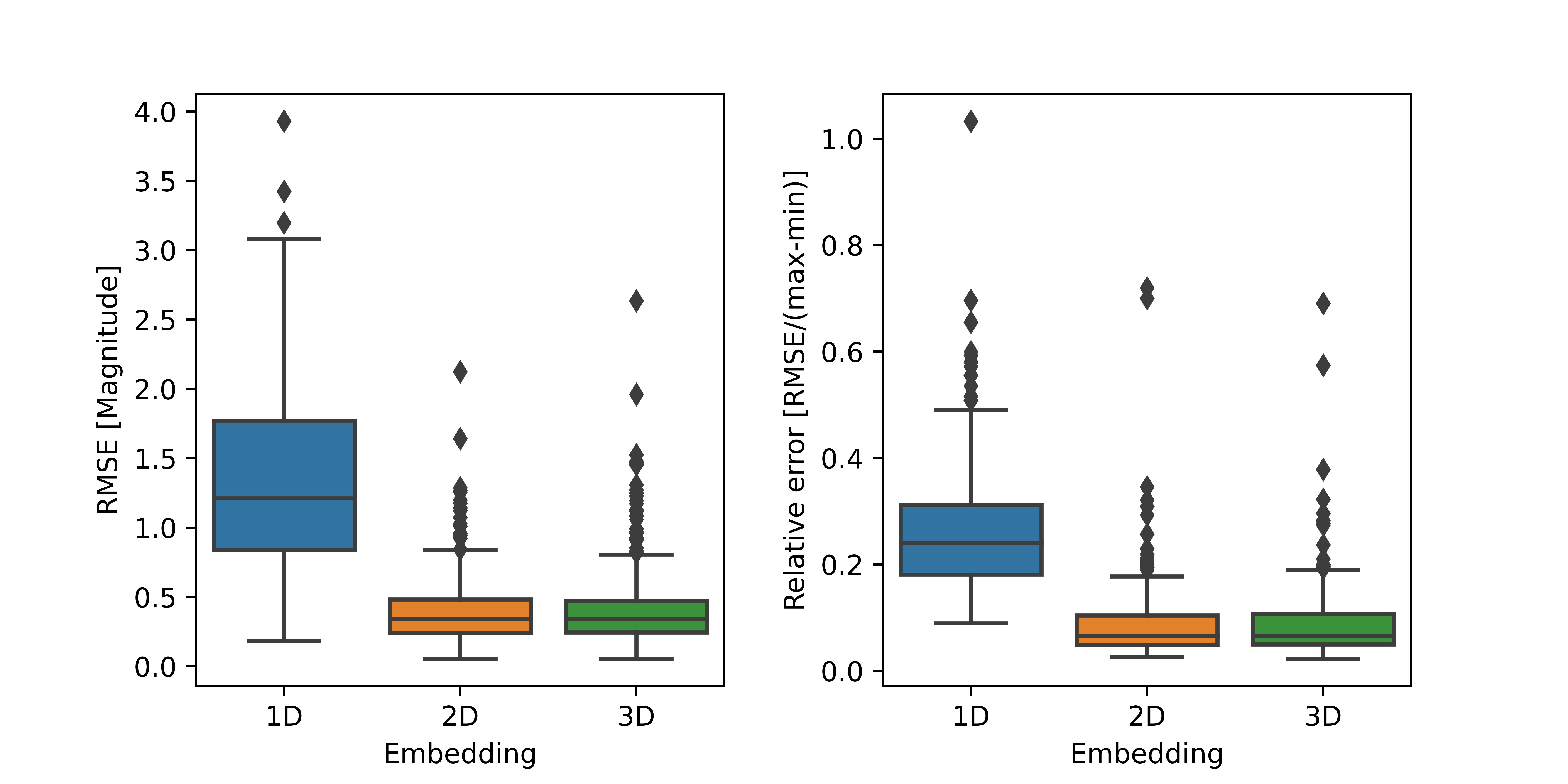}
  \caption{
    Accuracy of predictions for different data embeddings as a standard box
    plot representing the distribution of results of completing 5 random gaps
    of 4 distinct durations ranging from 3 hours to 2 weeks for each of the 13
    light curves in the dataset.
    1D is the raw time series of magnitudes, 2D takes into account daily
    periodicity, and 3D takes into account both daily and yearly periodicity.
    While GPs benefit greatly from the daily periodicity information, the
    effect of adding information about yearly periodicity is comparatively
    negligible.
  }\label{fig:results:embedings}
\end{figure}

\autoref{fig:results:embedings} shows the distribution of the resulting
comparison of the predictions to the truth for each embedding.
With a median of 0.066, the 2D embedding yields more than 3 times better
predictions than the original 1D embedding (median: 0.24).
The better prediction is possible because the distance between similar data
points across days is reduced, thanks to the extra dimension, allowing the GP
to more readily use them during interpolation.
However we see that adding a third dimension encoding the yearly periodicity
does not yield significant improvements over the 2D embedding since the results
are quasi-identical (median: 0.065).
Although the distributions of 2D and 3D embeddings are similar, from this on,
we focus on the 2D embedding and use it exclusively.

Part of the motivation for being able to predict data in light curves is to
complete gaps and missing data points.
A first benchmark is to be able to complete randomly selected data points
throughout a light curve.
We randomly select either 5\%, 10\%, or 20\% of the total number of data point
available in each of the 13 light curves in the dataset, 5 times for each
proportion.
Then, for more realistic interpolation, we select consecutive data points
representing gaps of either 3 hours, 1 day, 1 week, or 2 weeks, with 5
different random starting instants for each 13 light curve.
Lastly we repeat the same procedure but by selecting gaps at the end of each
light curve to evaluate how GPs perform for extrapolation.

\begin{figure}[htbp]
  \centering
  \includegraphics[width=\linewidth]{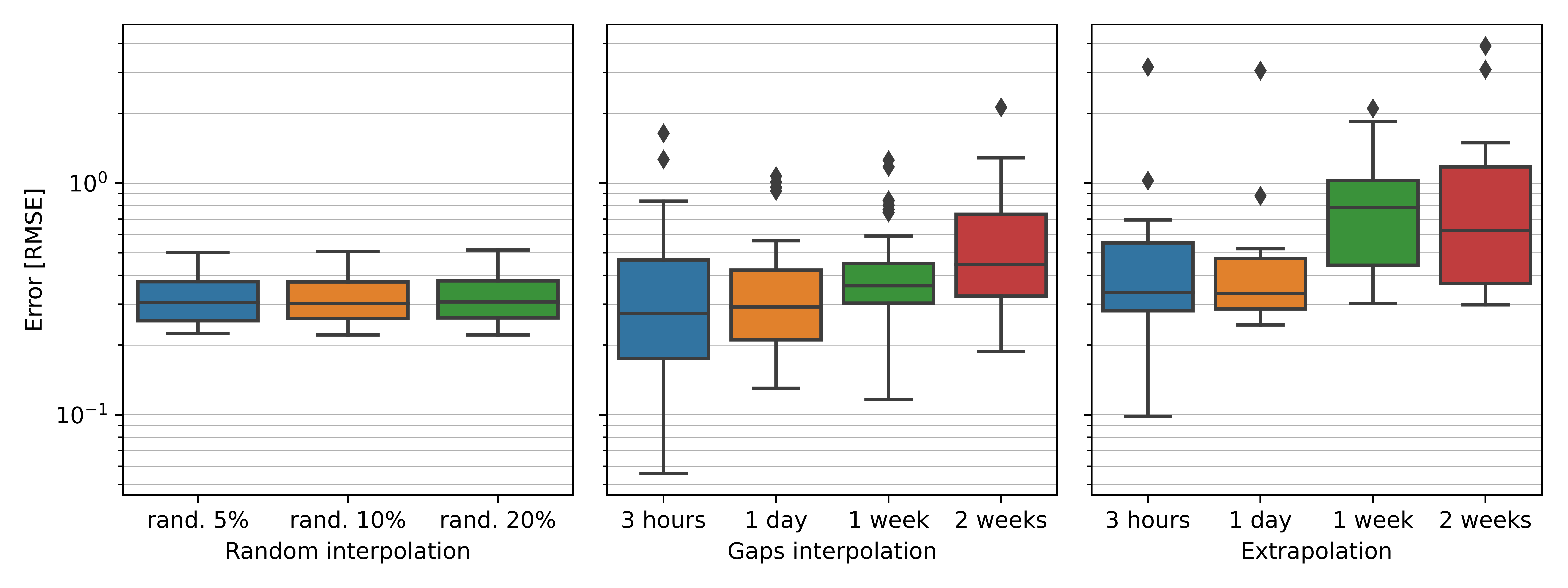}
  \caption{
    Comparison of the predicting power of GPs (with 2D embedding) for various
    interpolation and extrapolation tasks.
    The left pane is for interpolating randomly selected datapoint,
    representing from 5\% to 20\% of the entire light curves.
    The middle pane is for interpolating gaps of durations ranging from 3 hours
    to 2 weeks.
    The right pane is for extrapolating gaps of similar durations at the end of
    each light curves.
    Random data points are interpolated with great accuracy, even when they
    represent a significant portion of the data available.
    Interpolating gaps over short periods of time is more challenging.
  }\label{fig:results:interpolation_extrapolation}
\end{figure}

\autoref{fig:results:interpolation_extrapolation} shows the performance of GP
predictions for all of these interpolation and extrapolation tasks.
For random interpolation, for all percentages (5\%, 10\%, or 20\%), the
performance is similar and very good with a median relative RMSE of about 0.03.
For gaps, either interpolation or extrapolation, there is more variation, both
for a given gap duration, and across durations.

\begin{figure}[htbp]
  \centering
  \includegraphics[width=0.6\linewidth]{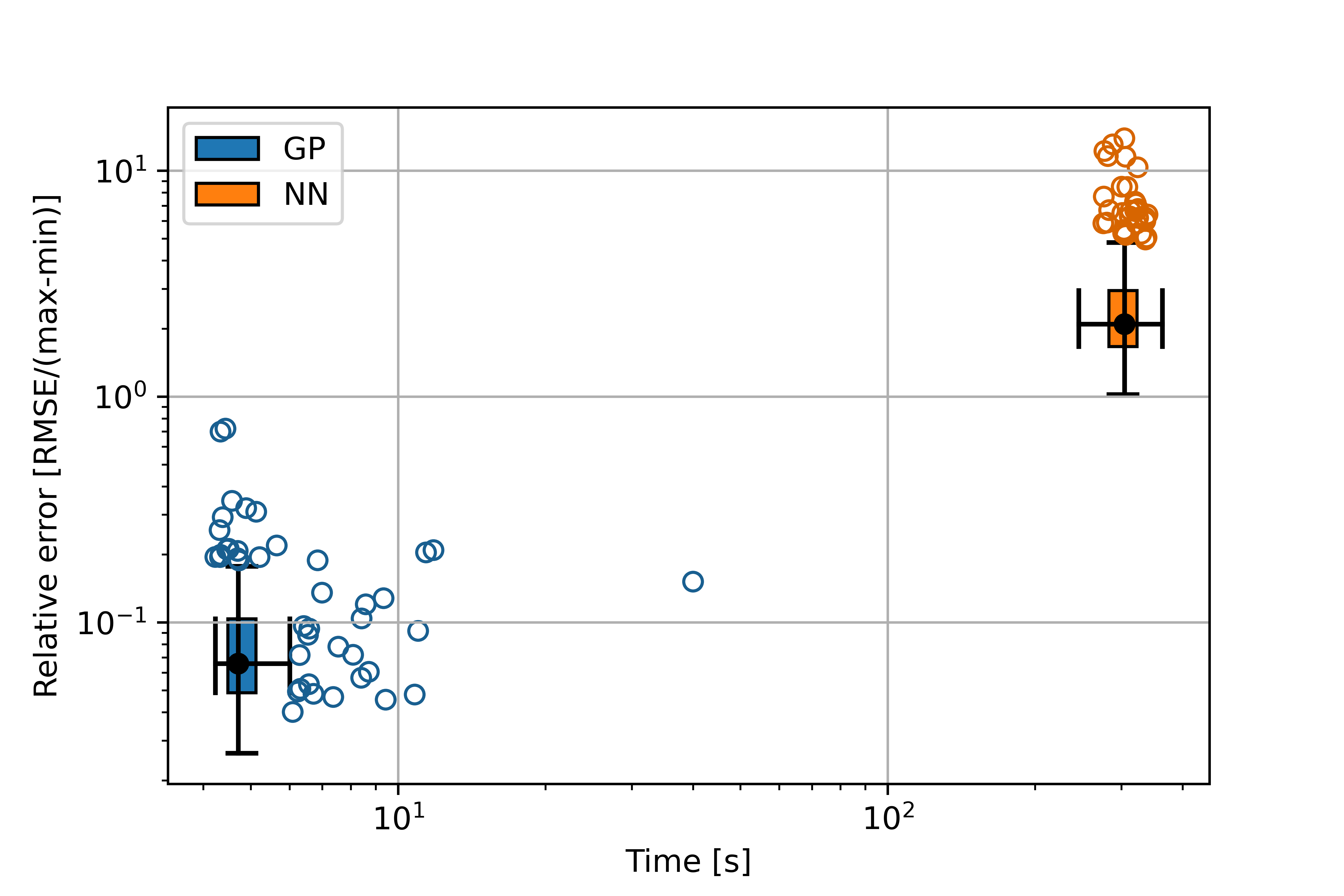}
  \caption{
    Comparison of the performance of our Gaussian process method with an
    equivalent neural network.
    The accuracy is measured by the RMSE scaled by the extent of the magnitude
    in the test dataset.
    The computing time is the total of the training and prediction time as
    those cannot be separated for the GP, and the prediction time is negligible
    compared to the training time for the NN.
    On the same data, the Gaussian process achieves a better accuracy in a
    fraction of the time.
  }\label{fig:results:nn_comparison}
\end{figure}

To compare the performance of GPs compared to DNNs, we ran the same task of
predicting one day of missing data in a single 4
year-long light curve with training data exclusively from that single light
curve on the same machine.
As shown in \autoref{fig:results:nn_comparison}, GPs achieve better accuracy in
only a fraction of the time.

\begin{figure}[htbp]
  \centering
  \includegraphics[width=0.6\linewidth]{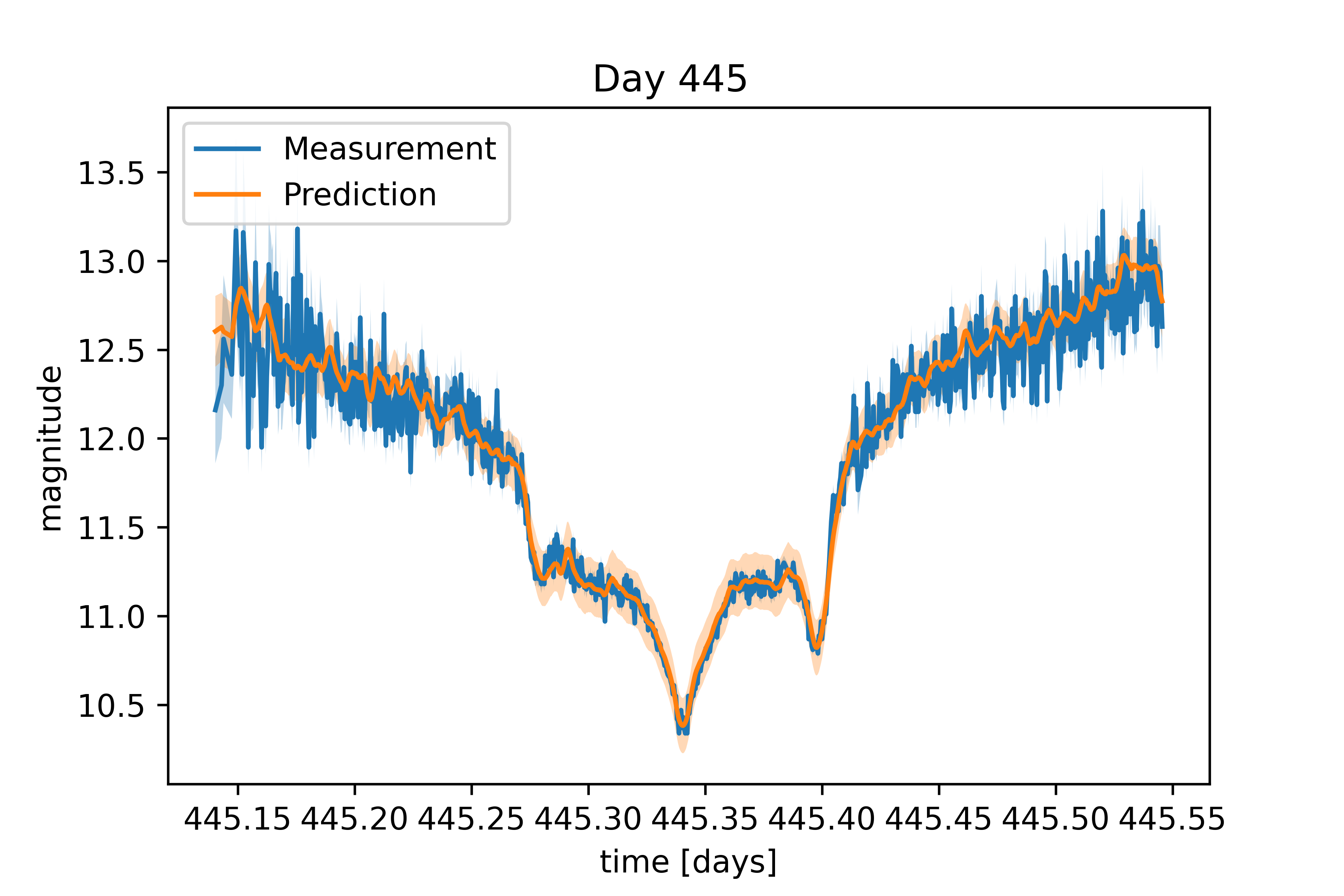}
  \caption{
    GP prediction (orange) compared to measurement (blue) for 1 day of missing
    data.
    The orange shaded region corresponds to the 95\% confidence interval of the
    prediction
    while the blue shaded region corresponds to the measurement standard error.
    This prediction was obtained using a RBF kernel with three dimensional
    training data.
    The GP does a very good job a predicting details while effectively
    filtering noise.
  }\label{fig:results:prediction}
\end{figure}

\autoref{fig:results:prediction} illustrates a single example of the quality of
the GP prediction
compared to raw measurements for a day of missing data using the 3D embedding.
The mean prediction follows the trend of magnitudes well as expected.
As is typical in GP models, our kernel model is assumed to be stationary and
homoscedastic, meaning that we learn a single set of hyperparameters that best
models all the training data (across all magnitudes).
In this interval pictured, we see that the data at high magnitudes seem to have
more variance than the data collected at lower magnitudes.
Our model is agnostic to any change in variance in the data itself, but will on
average provide desired uncertainty quantification.
Note this one timegap is a single interval that is in itself time-correlated in
magnitude and therefore variance regime.
Therefore, the coverage of the 95\% interval from this one sample can be
different than the desired overall level, but when we consider many independent
time intervals of this type, the uncertainties will average to the desired 95\%
confidence level.
In future work more flexible and complex GP kernel functions could be designed
with inhomogeneous or non-stationary kernels to correct for this potential
pattern in magnitude to improve uncertainties in each individual sample region.
If the magnitude distribution in our prediction interval differs from that of
our training data then there is a model discrepancy in our approach that could
explain our observed variances.

\begin{figure}[htbp]
  \centering

  \includegraphics[width=0.5\linewidth]{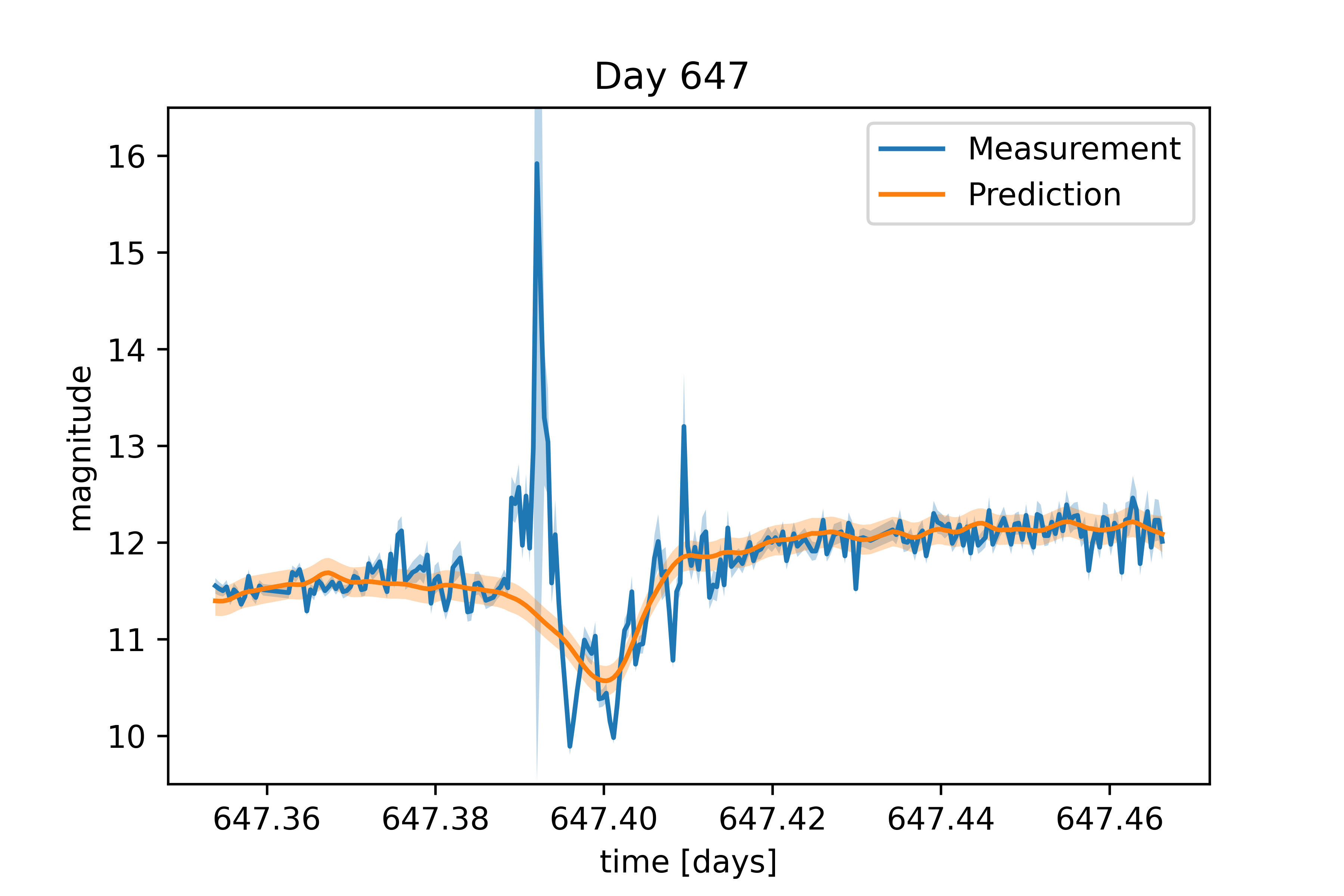}%
  \includegraphics[width=0.5\linewidth]{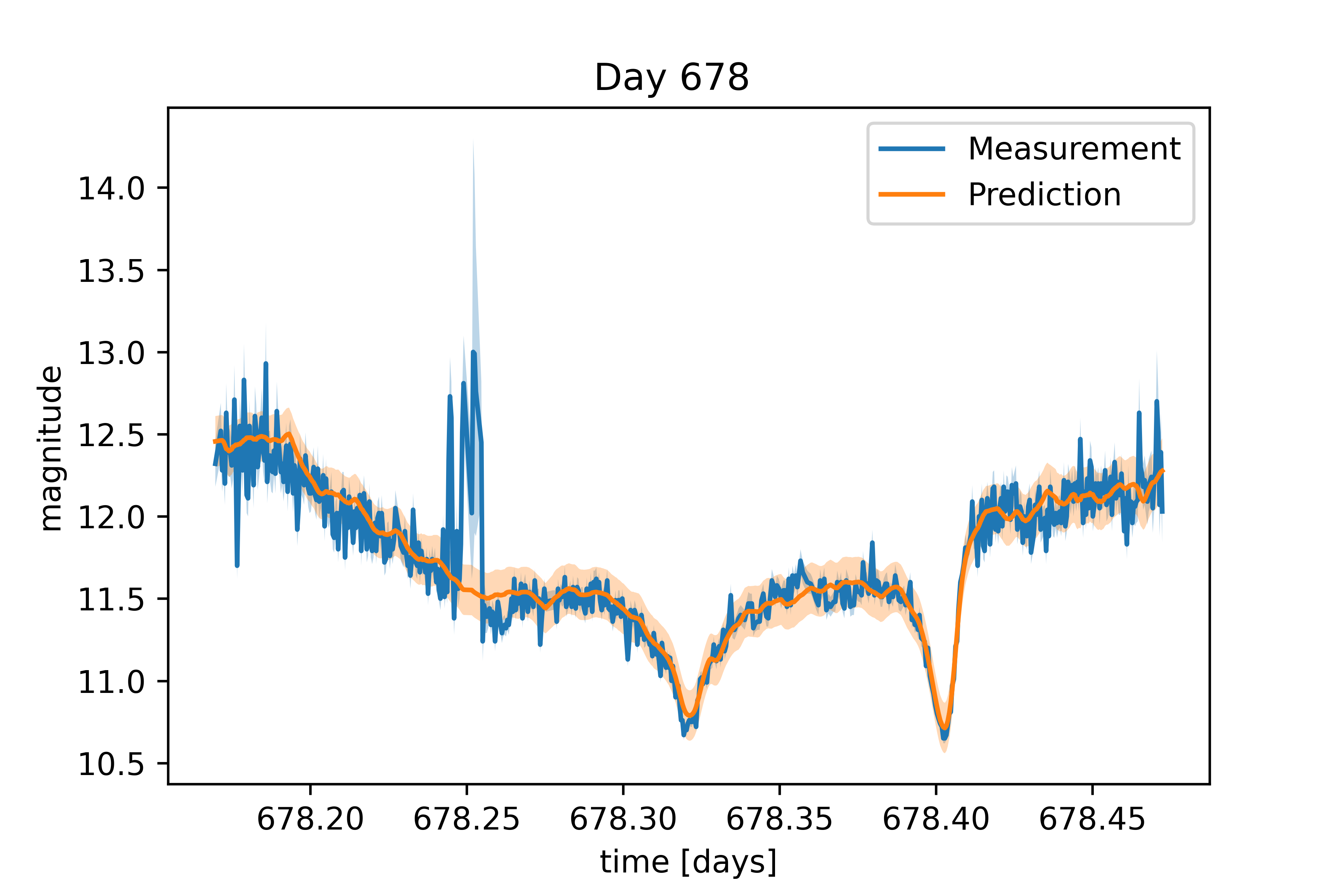}
  \caption{
    Example of detected anomalies.
    The prediction is smooth and matches the measurement well except for a
    short time interval (near 647.39 and 678.25 days respectively) where the
    measurement suddenly exceeds its expected value based on the prediction and
    confirmed by measurements from the surrounding days.
  }\label{fig:results:anomalies}
\end{figure}

But the true power of GPs comes from their ability to predict a full posterior
distribution and not just a mean.
The predicted covariance can be used to detect areas where the prediction
differs from the measurement by a significant amount.
One application of particular interest for SDA is the
detection of anomalies --- potential maneuvers or state changes --- in light
curve data.
\autoref{fig:results:anomalies} shows two example of such anomalies.
In both cases, two factors make the anomalies noticeable.
Firstly, the prediction differs from the measurement for a small but
significant period of time, and the difference exceeds two times the predicted
standard deviation.
And secondly, the measurements from the surrounding days look similar to those
of the focal day --- and to the prediction --- except for that particular
period.
The combination of those two factors indicate that it is indeed the
measurements that are anomalous and not the predictions.

\section{Conclusion}
\label{sec:conclusion}

We have shown Gaussian processes (GPs) using MuyGP to be a capable and
pertinent tool for analyzing light curve data.
Unlike other machine learning methods that can have millions of parameters,
Gaussian processes are simple in that they do not have a large architecture
search and in that they only have a few parameters to be estimated.
Unlike traditionally GP estimation methods, the implementation of the MuyGPs
methodology allows astronomers to apply GP methods to very large data with very
little computational or memory burden.
In addition to their improved accessibility, we have shown that in our designed
experiments on light curves, Gaussian processes are both more accurate and
faster than an example neural network.
Further, the uncertainty quantification we get from Gaussian processes allows
us to identify statistically distinct deviations from the pattern-of-life.


As seen above, while Gaussian processes are able to ingest raw uni-dimensional
time series, encoding the periodicity of the light curve data through
additional dimensions yields much better predictions when we predict large gaps
of observations.
This is likely due to the distinct periodic trends in the light curves.
However, we observed that encoding the daily periodicity was sufficient and
that encoding the yearly periodicity did not further improve prediction.
A first explanation is that surrounding days are typically more similar than
surrounding years, so the majority of ``nearest neighbors" used in the GP model
would naturally be selected from surrounding days rather than from surrounding
years.


We have also shown how Gaussian processes compare advantageously to neural
networks both in terms of computing time during training and prediction
accuracy.
Note that our metric for computing time combines both training and prediction
time
from the two models.
This metric masks the tradeoff that a trained neural network can make
additional predictions efficiently whereas training a Gaussian process using
MuyGPs is very fast, but the prediction time given a trained model is much
slower comparatively.
\citep{fastmuygps} demonstrate a way to improve prediction time
of a simlar kernel interpolation with an additional approximation.
However, in the online scenario of the light curve problem where one may want
to perform this analysis in near realtime, one would both retrain and predict
simultaneously so the GP method is significantly preferred.

Both methods can be parallelized to take advantage of High Performance
Computing (HPC).
For DNNs, parallelizations techniques are readily available, for instance in
PyTorch, with the advent of GPUs and TPUs.
For MuyGPs, parallelizations efforts are in progress and will be released in a
different publication.
We ran all of the experiments in this manuscript on a single core of a
commodity laptop using the software library MuyGPyS~\cite{muygpys2021github}.


Future applications will use MPI~\cite{dalcin2011parallel} and
JAX~\cite{jax2018github} to scale model training and evaluation to multiple
compute nodes using many CPUs and GPUs in parallel.
These scalability features will enable applications that scale to the
observation sizes of LSST, which are anticipated to involve hundreds of
millions of space objects.



Possible future directions of research involve comparing GP training
from one set of light curves and using these trained parameters to make
predictions for a
different set of unseen light curves.
This will determine whether it is necessary to fit parameters to each light
curve.
Finally, since our method gives fully interpolated light curves, future work
could use these de-noised full time series for further SDA studies.

In summary, we believe our GP method is a valuable method for light curve
interpolation
and future prediction for three scientific tasks of particular interest.
First, by de-noising and interpolating missing data, GPs can be part of a
pre-processing pipeline before feeding the data to further algorithms or
software that cannot work with raw data.
As shown in figures~\ref{fig:illustration:light_curve}
and~\ref{fig:illustration:missing-data}, raw light curves can have a lot of
gaps frequently spanning multiple days so being able to interpolate those gaps
is crucial.
Second, by being able to extrapolate data over several hours, days, or even
weeks, GPs can be used to predict the expected behavior of RSOs, which can
serve in forecasting and collision avoidance in SDA.
Last and not least, by quantifying the uncertainty of predictions, GPs allow
for automatic change detection.
Indeed, being able to predict a full posterior distribution enables discerning
anomalies in measurements from anomalies in prediction, since uncertain
predictions can easily lead to false positives.
In future research, it should be possible to use tracking data and/or
documented known maneuvers to fine-tune and benchmark our maneuver detection
capabilities.

\section*{Acknowledgments}

This work was performed under the auspices of the U.S.
Department of Energy by
Lawrence Livermore National Laboratory under Contract DE-AC52-07NA27344 with IM
release number LLNL-PROC-839253.

Funding for this work was provided by LLNL Laboratory Directed Research and
Development grant 22-ERD-028.

This document was prepared as an account of work sponsored by an agency of the
United States government.
Neither the United States government nor Lawrence
Livermore National Security, LLC, nor any of their employees makes any
warranty, expressed or implied, or assumes any legal liability or
responsibility for the accuracy, completeness, or usefulness of any
information, apparatus, product, or process disclosed, or represents that its
use would not infringe privately owned rights.
Reference herein to any specific
commercial product, process, or service by trade name, trademark, manufacturer,
or otherwise does not necessarily constitute or imply its endorsement,
recommendation, or favoring by the United States government or Lawrence
Livermore National Security, LLC.
The views and opinions of authors expressed
herein do not necessarily state or reflect those of the United States
government or Lawrence Livermore National Security, LLC, and shall not be used
for advertising or product endorsement	purposes.

\bibliographystyle{plainnat}
\bibliography{AMOS_2022}

\end{document}